# THz-frequency cavity magnon-phonon-polaritons in the strong coupling regime

Prasahnt Sivarajah, Andreas Steinbacher, Blake Dastrup, and Keith Nelson

*Massachusetts Institute of Technology, 77 Massachusetts Avenue, Cambridge, MA, USA 02139*

(Dated: July 11, 2017)

We demonstrate the strong coupling of both magnons and phonons to terahertz (THz) frequency electromagnetic (EM) waves confined to a photonic crystal (PhC) cavity. Our cavity consists of a two-dimensional array of air-holes cut into a hybrid slab of ferroelectric lithium niobate (LiNbO$_3$) and erbium orthoferrite (ErFeO$_3$), a canted antiferromagnetic crystal. The phonons in LiNbO$_3$ and the magnons in ErFeO$_3$ are strongly coupled to the electric and magnetic field components of the confined EM wave, respectively. This leads to the formation of new cavity magnon-phonon-polariton modes, which we experimentally observe as a normal-mode splitting in the frequency spectrum and an avoided crossing in the temperature-frequency plot. The cavity also has a mode volume of $V = 3.4 \times 10^{-3}\lambda^3 \simeq 0.5(\lambda/n)^3$ $\mu m^3$ and can achieve a Q-factor as high as 1000. These factors facilitate the pursuit of the fields of THz cavity spintronics and quantum electrodynamics.

## I. INTRODUCTION

The light-matter interaction between an electromagnetic field and a resonant material mode can be enhanced when it occurs inside of a cavity that strongly confines the light. The interaction inside the cavity is termed as strong coupling (SC) if the rate of energy transfer between the light and matter is much faster than the irreversible processes due to both loss of light out of the cavity and non-radiative damping within matter [1]. In the SC regime, light and matter can no longer be treated as separate entities and must be described by two superposition states called cavity polaritons with an energy difference given by the normal mode or Rabi splitting energy $\hbar\Omega$. Cavity-enhanced SC has been observed between photons and a diverse range of excitations including Rydberg atom transitions [2], excitons [3], plasmons [4], NV center spins [5], magnons [6], and Cooper pairs of a superconductor [7]. A subset of demonstrations have been at the single-photon or few-photon level, and these fall under the field of cavity quantum electrodynamics (QED) [8, 9]. However, strong coupling can also be based on the classical coherence of macroscopic electrodynamic fields. While the QED demonstrations motivate cutting-edge applications such as quantum computation [10], classical demonstrations are also valuable. Not only do they serve as important groundwork for their quantum analogues, but they also have applications in their own right in fields such as lasing [11, 12] and sensing [13, 14].

The value of classical demonstrations can be seen in the field of cavity spintronics, which studies the strong coupling of cavity photons to spin ensembles (e.g. magnons). The first classical demonstration of cavity spintronics was reported by Zhang et al. [15] using microwave cavity photons and magnons in yttrium iron garnet (YIG). They observed a Rabi-like oscillation of the light intensity, which they attributed to the coupling between the magnetic field of the EM wave and the magnetization of YIG. Thus far, cavity spintronics has demonstrated the ability to achieve long-term storage and all-optical addressing of spin states, important elements in realizing spin-based computation.

More capabilities can be added to cavity spintronics by transitioning it into the THz frequency range. The first report of polaritons formed by the SC of THz light and magnons was by Sanders et al. in 1978 [16], wherein a slab of iron (II) fluoride that exhibits an antiferromagnetic resonance at 1.58 THz was used. More recently, Kampfrath, et al. used advancements in the generation of ultrashort and intense THz pulses to demonstrate coherent control over antiferromagnetic magnons in nickel oxide [17]. There have also been demonstrations at THz frequencies of the inverse Faraday effect [18] and nonlinear spin control [19, 20]. The prospect of harnessing these properties for cavity spintronics makes it desirable to realize a platform at THz frequencies. However, such a platform has remained elusive so far, mostly due to the absence of highly confining THz cavities and sensitive detection techniques.

Here, we overcome both of these limitations by using a unique photonic crystal (PhC) cavity. We demonstrate strong coupling of both magnons and phonons to THz cavity photons. Specifically, we build a PhC cavity in hybrid structure consisting of a 53 $\mu m$ slab of lithium niobate (LiNbO$_3$) and a 40 $\mu m$ slab of erbium orthoferrite (ErFeO$_3$). Optical phonons in the slab of ferroelectric LiNbO$_3$ are strongly coupled to a THz E-field to form phonon-polaritons [21, 22], while the magnons in an adjacent slab of the canted antiferromagnetic ErFeO$_3$ are strongly coupled to the THz magnetic field (H-field) to form magnon-polaritons [23]. We recently demonstrated coupling of the two modes in a hybrid LiNbO$_3$/ErFeO$_3$ waveguide to form new hybridized modes that we termed magnon-phonon-polaritons [24]. In the present work we have extended the demonstration to a hybrid 3D cavity through spatial patterning of the hybrid waveguide structure. In the hybrid cavity, the bare magnon, phonon, and photon modes become strongly coupled to form new hybridized modes that we term cavity magnon-phonon-polaritons. Coupling to the magnons in ErFeO$_3$ forms the basis for a spintronics platform, while coupling to the phonons in LiNbO$_3$ enables the efficient generation and detection of light directly inside the cavity. In Sec.

II of this article, we will first examine cavity phonon-polaritons using a PhC cavity fabricated in a slab of LiNbO$_3$, and in doing so point out the unique properties of the platform. In Sec. III, we will then examine cavity magnon-phonon-polaritons using a PhC cavity fabricated in a hybrid slab consisting of both LiNbO$_3$ and ErFeO$_3$. We will validate our observation of strong light-matter coupling by demonstrating normal-mode splitting in the THz frequency spectrum and avoided crossing in the temperature-frequency dependence.

## II. PHONON-POLARITONS IN A PHC CAVITY

Initially, the sample under investigation is a 53 $\mu$m thick (100) slab of LiNbO$_3$. By focusing an intense 800 nm laser pulse into the slab, coherent THz phonon-polariton wavepackets can be generated via impulsive stimulated Raman scattering (ISRS) [25–28], a process that is exploited extensively for high-field THz wave generation in LiNbO$_3$ [29]. Our phonon-polariton wavepackets, with ∼0.1-2 THz spectral content, are in the lower polariton branch that extends below the optical phonon frequency of 7.4 THz and corresponds to in-phase excursions of the lattice vibration and the EM field. Due to the index mismatch between the phonon-polaritons and the optical pulse generating them, the phonon-polaritons propagate away from the generation region. In a bulk LiNbO$_3$ crystal the THz wave propagation is primarily in the lateral directions relative to the optical pump beam propagation direction, and in a LiNbO$_3$ slab whose thickness is on the order of the THz wavelength, the THz waves propagate in the plane of the slab as dielectric waveguide modes [30]. To detect their progress, we exploit the electro-optic (EO) effect wherein the THz electric field (E-field, $E_{THz}$) modulates the LiNbO$_3$ refractive index [31]. More specifically, the lattice displacements of the optical phonon mode change the optical refractive index via electron-lattice interactions [32]. By passing a time-delayed optical probe pulse through the slab and measuring the THz-induced depolarization (see Fig. 1a), a time-dependent THz electric-field profile $E_{THz}(t)$ can be obtained (see Fig. 1b). The field profile can then be Fourier transformed (FT) to retrieve the frequency spectrum (see Fig. 1c). If the probe pulse is spatially expanded and detected on a camera, we can also obtain a spatially resolved image of the THz wave (see Fig. 1d). We thus have a platform for generating and detecting THz waves directly inside the slab. In the past, these basic properties were used to demonstrate capabilities such as spatiotemporal coherent control of lattice vibrational waves [33], time-resolved near-field THz imaging [34], and on-chip all-dielectric devices [35]. Note that although the THz waves in our experiments are largely light-like, the phononic nature is still integral to the dynamics. Not only does the lattice carry about 30% of the energy in the 0.1-2 THz range [21] (see Appendix), the phonon mode is also responsible mechanistically for effi-cient generation and detection of the THz waves in our experiments. In addition, we will see shortly that the phononic nature also plays the dominant role in defining the resonant frequency and lifetimes of the cavity modes.

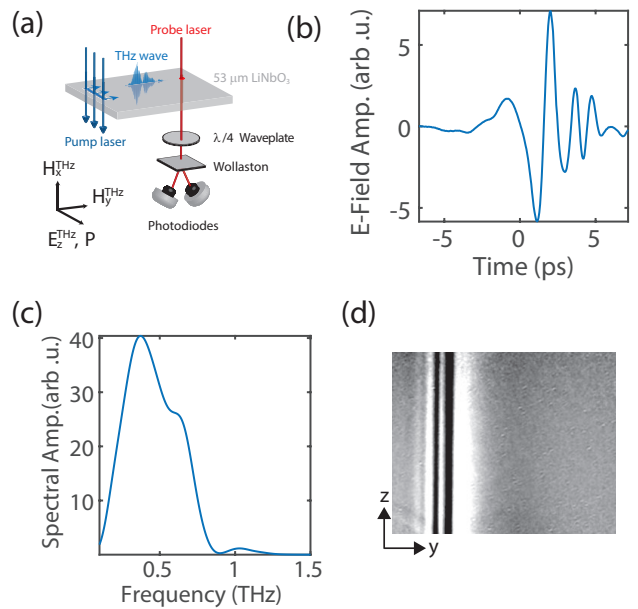

FIG. 1. **On-chip THz generation and detection.** (a) Schematic illustration of the use of a line-focused 400 nm pump pulse to generate THz frequency phonon-polaritons in a thin slab of LiNbO$_3$ (30-100 $\mu$m thick). An 800 nm probe pulse is variably delayed from the pump pulse to measure the time-dependent THz E-field profile via the EO effect. The coordinate axes represent the polarization of the THz light and the LiNbO$_3$ ferroelectric polarization $P$ that is modulated by the phonon-polariton wave. (b) Time-domain profile of the detected THz wavepacket and (c) the corresponding THz spectral intensity, obtained by squaring the FT of (b). (d) Recorded image of a THz wave propagating inside a LiNbO$_3$ slab, obtained from the difference between signals on a camera measuring light intensities in orthogonal polarization as shown in (a). The signal at any pixel in the image is directly proportional to the THz E-field, with the white and black regions representing opposite polarities of the E-field.

The bare LiNbO$_3$ waveguide provides THz field confinement along the $x$-axis (out-of-plane axis), but to build a cavity we also desire confinement in the $y$- and $z$-directions. To accomplish this, we utilized a photonic crystal cavity [36–38] composed of a periodic array of air holes cut into the LiNbO$_3$ slab using femtosecond laser machining [39]. The cavity we fabricated, often referred to as an L3 cavity [40], is shown in Fig. 2a and is composed of a hexagonal lattice of air holes with three holes removed from the array to form a "defect". To excite the cavity modes, we focused a 400 nm pump pulse into the defect region of the crystal, which launched two counter-propagating THz phonon-polariton wavepackets. To detect these cavity modes, we directed an 800 nm probe pulse into the defect region and recorded the time-evolution of the THz E-field as shown in Figs. 2b (imaging) and 2c (localized point detection). The wavepackets initially contain a broad range of frequencies as evidenced by a windowed FT (see Fig. 2c, right). When they reach the edges of the cavity, those waves with frequencies in the PhC band gap are strongly reflected while the rest of the frequency components are partially reflected due to the index mismatch with the surrounding PhC (i.e. Fresnel reflection). The bandwidth in the cavity thus narrows with every successive interaction with the holes, quickly approaching the limit where only the cavity modes are confined to the slab (see Fig. 2c, right). In Fig. 3a, we show the simulated E-field profiles for two of these modes, labelled X1 and X2. In general, the modes have distinct spatial distributions, with the higher-frequency modes exhibiting a higher number of nodes and extrema. Due to the small focal spot size of the pump ($w_y \sim 10$ $\mu m$, $w_z \sim 50$ $\mu m$) and probe pulses ($w_y \sim w_z \sim 10$ $\mu m$), their locations in the cavity determine the modes with which they predominantly interact. By choosing pump and probe pulses to overlap with the extrema of a particular mode spatially, we can favor its generation and detection, respectively. For instance, in Fig. 3b we show the frequency spectrum where we selectively excited and detected either the X1 or X2 mode.

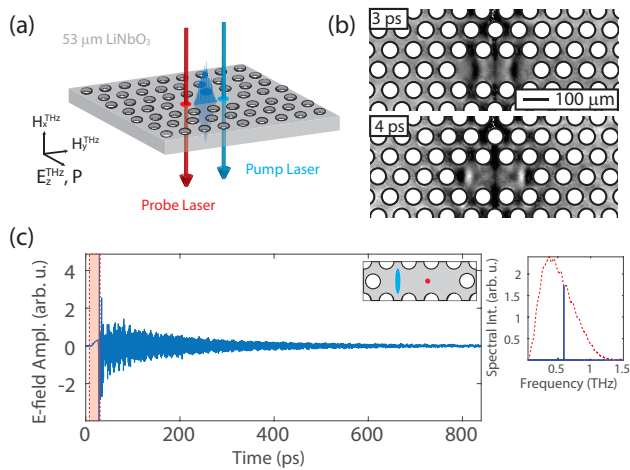

FIG. 2. **THz-frequency phonon-polaritons in a PhC cavity.** (a) Schematic illustration of a L3 PhC cavity composed of a hexagonal lattice of air holes cut through a 53 $\mu m$ LiNbO₃ slab, with three air holes removed from the array to form the defect. Optical laser pulses are focused into the defect region to generate and detect phonon-polariton cavity modes. The coordinate axes represent the polarization of the THz light and the LiNbO₃ ferroelectric polarization $P$. (b) Recorded images at 3 and 4 ps after generation, showing THz phonon-polaritons counter-propagating away from a pump pulse that irradiated the center of the PhC cavity (air holes: radius $r = 29$ $\mu m$, periodicity $a = 100$ $\mu m$). (c) The left figure is a time trace of the THz E-field in the cavity, taken by fixing the location of the pump and probe lasers as shown in the inset. Dashed lines indicate an approximate time window during which the initial THz wavepacket passes by the probe pulse. The right figure shows the spectral intensity of the windowed region (red dashed lines) and the entire time trace (blue solid line), demonstrating that the initial broad bandwidth is narrowed to that of the cavity mode.

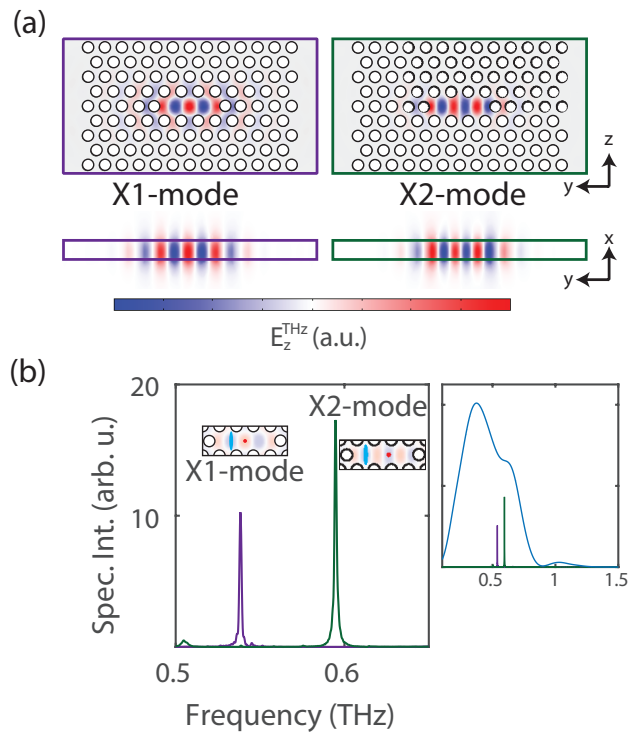

FIG. 3. **Selective excitation and detection of cavity modes.** (a) FEM simulations of the mode profiles for two PhC cavity modes (air holes: radius $r = 29$ $\mu m$, periodicity $a = 100$ $\mu m$) denoted as X1 and X2. The color map corresponds to the $z$-polarized electric-field distribution (b) The left figure shows the frequency spectrum after selective interaction with either the X1 or X2 mode, with accompanying insets that show the location of pump (blue) and probe (red) beams. The right figure compares the narrow linewidths of the cavity modes to the broad spectrum in the bare waveguide.

Thus far, we have discussed the central role that the phonon mode plays in THz wave generation and detection, but the phononic nature of the polaritons is most easily seen by its influence on the cavity modes as a function of temperature. Firstly, the transverse optical phonon in LiNbO₃ is a soft mode associated with the ferroelectric phase transition at $T_c = 1418$ K [41] at which the dielectric constant reaches a maximum. As a result, there is a sizeable decrease in the THz refractive index from its room temperature values ($n_{eo} \sim 5$ & $n_o \sim 6.7$) to those at cryogenic temperatures [42]. As shown in Fig. 4, the decrease in the refractive index increases the cavity mode frequency ($\Delta\omega/\omega \sim \Delta n/n$ [38]). Secondly, the damping of the low-frequency phonon-polaritons in LiNbO₃ is mainly attributed to a coupling of the TO phonon to other low-frequency phonon modes [43, 44].

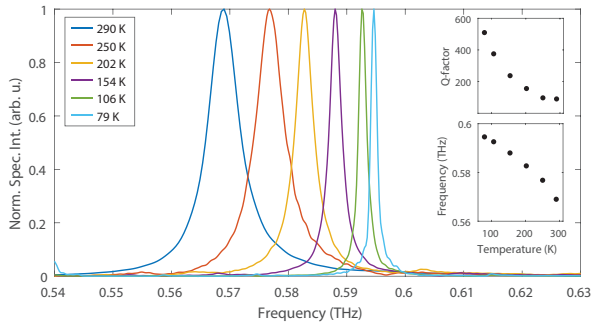

FIG. 4. **Temperature-dependent frequency spectra.** The temperature-dependent spectra for the X2 cavity mode (see Fig. 3a) demonstrate the influence of the soft transverse optical phonon mode in LiNbO₃. As the temperature is reduced, the refractive index and phonon damping decrease and this causes, respectively, an increase of the mode frequency and a decrease of the mode linewidth. Inset shows the temperature dependence of the Q-factor and resonant frequency.

This damping is reduced by a factor of ∼5 from 290 K to 79 K as the low-frequency phonon modes become depopulated [42, 45]. The effect on the cavity modes can be deduced from Fig. 4, where the mode linewidth is also reduced by a factor of ∼5. To analyze this change more quantitatively, we introduce a measure of the cavity lifetime called the quality factor $Q = \omega_{cav}/\kappa$, where $\omega_{cav}$ is the resonant frequency of the mode and $1/\kappa$ is the rate of energy loss from the cavity [46]. The quality factor can be interpreted as a dimensionless lifetime that can account for the combined contributions from the intrinsic damping $Q_m$ and radiative scattering of light out of the cavity $Q_r$ leading to a net $Q_{net}$ defined as $1/Q_{net} = 1/Q_m + 1/Q_r$. In our case, the increase in $Q_{net}$ with temperature is directly proportional to the decrease in phonon damping because that is the dominant mechanism limiting the lifetime, i.e. $Q_m \ll Q_r$ so $Q_{net} \simeq Q_m$. Hence, in our experiments the highest $Q = 510$ was obtained at our lowest recorded temperature of 79 K. Based on the temperature-dependence of the phonon absorption coefficient of LiNbO₃, the $Q$ can be increased further to $Q \simeq 1000$ at 4K [45]. Another important measure of the cavity is the effective mode volume $V_{eff}$, defined as

$$V_{eff} = \frac{\int \epsilon(\mathbf{r}) \left| \mathbf{E}(\mathbf{r}) \right|^2 d^3 \mathbf{r}}{\max \left[ \epsilon(\mathbf{r}) \left| \mathbf{E}(\mathbf{r}) \right|^2 \right]} \qquad (1)$$

For the X1 mode, $V_{eff} = 3.4 \times 10^{-3} \lambda^3 \simeq 0.5(\lambda/n)^3$, where $\lambda$ is the free-space wavelength and $n = 5.21$ is the average refractive index of the slab. Although the effective mode volume is not appreciably smaller than the $(\lambda/n)^3$, the large refractive index of the LN slab at THz frequencies results in a mode volume that is nearly three orders of magnitude smaller than the free-space volume.

## III. MAGNON-PHONON POLARITONS IN A PHC CAVITY

Having established the particulars of our platform, we now discuss our experiments regarding the strong coupling between magnons, phonons, and cavity photons. Our system consists of a thin composite slab of 53 μm thick (100) LiNbO₃ and 40 μm thick (001) ErFeO₃. In LiNbO₃, we have seen that the material excitation is the polar lattice vibration with its polarization $P$ along the $z$-axis (as defined in Fig. 5a). In ErFeO₃, the material excitation is the collective magnetic spin, i.e. the Brillouin zone center quasi-antiferromagnetic (AF) magnon mode. The THz H-field modulates the net magnetization $M_{AF}$ along the $c$ axis of the orthorhombic crystal ($x$-axis as defined in Fig. 5a) in amplitude at the magnetic resonance transition frequency (0.67 THz at 290 K, 0.75 THz at 79 K). As before, a PhC cavity was fabricated using laser machining to cut air-holes in a hexagonal array through the hybrid slab consisting of LN and EFO (see Fig. 5a). The simulated $E_z$-field and $H_x$-field profiles for the cavity modes (labelled X1) are shown in Fig. 5b. The E-field component of the light, polarized along the $z$-direction, can couple to the phonons in LiNbO₃ to produce mixed phonon-polariton modes. Simultaneously, the H-field component of light, polarized along the $x$-axis, can couple to the AF mode of EFO to produce mixed magnon-polaritons. In the hybrid cavity, uniform $z$-electric-polarized, $x$-magnetic-polarized THz-frequency electromagnetic waves can extend throughout both materials, which enables us to form new hybridized cavity magnon-phonon-polariton modes.

To verify the formation of cavity magnon-phonon-polariton modes, we studied the frequency spectrum of the THz waves as the frequencies of the cavity and magnon modes were tuned by temperature. While we have seen that the cavity resonance of the pure LN shifts upon reduction of the temperature from 290 K to 77 K by ∼18 GHz in a weakly quadratic fashion (see Fig. 4 inset), the magnon resonance shifts by ∼80 GHz and does so with a higher-order functional form [47, 48]. To record the frequency spectrum, we generated and detected THz waves through ISRS and EO detection as described earlier, in the LiNbO₃ portion of the hybrid cavity (see Fig. 6a). Despite being excited in LiNbO₃, the THz waves evolve into the modes that span the entire hybrid structure. Thus, there exists an E-field parallel to the polar lattice vibration in LiNbO₃ and an H-field parallel to the antiferromagnetic magnetization in ErFeO₃. However, EO probing is only sensitive to the E-field in LiNbO₃ and can therefore only detect the AF-mode magnon response through its interaction with the cavity mode. As shown in Fig. 6b, at 290 K we only observed a single Lorentzian peak corresponding to the cavity mode at 0.72 THz and did not observe the magnon mode at 0.67 THz, which was sufficiently detuned such that any interaction with the cavity mode was negligible. However, as the temperature was reduced and thus the two modes tuned to be in resonance, a clear double-peaked spectrum appeared. At 210 K, the detuning between the bare modes was only 2 GHz, yet we observed two peaks that were separated by nearly 8 times the detuning. This double-peaked structure is characteristic of a normal mode splitting that can arise from the strong coupling between the cavity and magnon modes.

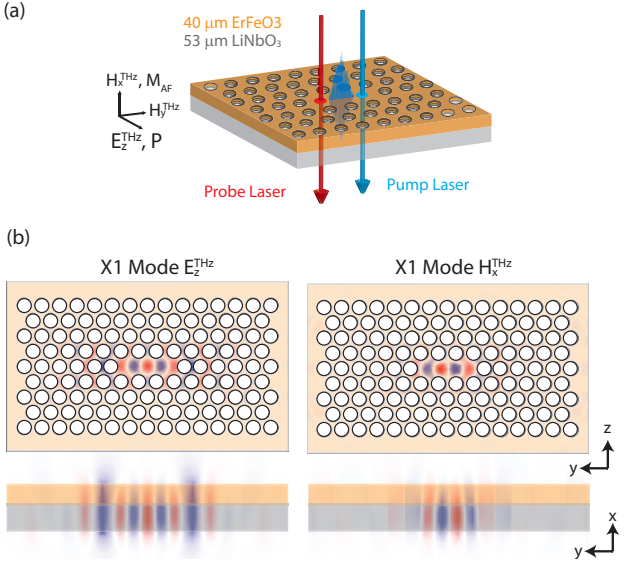

(a)

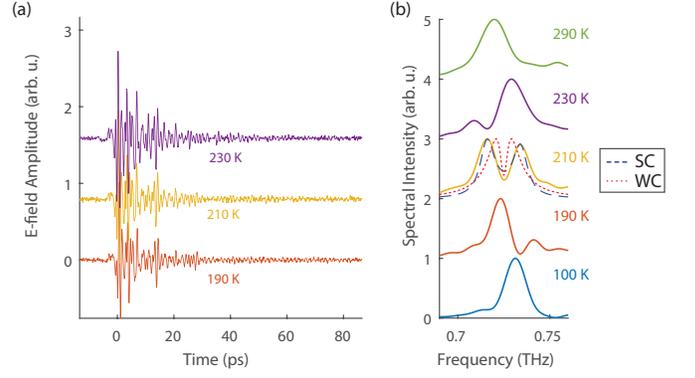

FIG. 5. **Hybrid PhC cavity.** (a) Schematic illustration of a L3 PhC cavity composed of a hexagonal lattice of air holes cut through a hybrid slab composed of 53 $\mu$m LiNbO$_3$ and 40 $\mu$m ErFeO$_3$. Optical laser pulses are focused into the defect region to generate and detect magnon-phonon-polariton cavity modes. The coordinate axes represent the polarization of the THz light, the LiNbO$_3$ ferroelectric polarization $P$ that is modulated by the phonon-polariton mode, and the ErFeO$_3$ magnetization $M_{AF}$ that is modulated by the AF magnon mode. (b) FEM simulation of the X1 mode profile in the hybrid PhC cavity (air holes: radius $r = 29$ $\mu$m, periodicity $a = 74$ $\mu$m). The left image shows the $y$-polarized electric-field distribution, and the right image shows the $z$-polarized magnetic-field distribution.

FIG. 6. **THz-frequency magnon-phonon-polaritons in a PhC cavity.** (a) Temperature-dependent time traces of the THz E-field in the hybrid cavity. Traces are vertically offset by 1 unit for clarity. (b) THz spectra near the X1 mode. The double-peaked spectrum in the range of 170-230 K demonstrates the presence of a normal mode splitting between the cavity and magnon modes. The dashed and dotted lines overlaid at 210 K correspond to models for strongly coupled oscillators (SC) and THz absorption by the magnon mode in the weak-coupling limit (WC), respectively.

The interaction can be modelled using the Hamiltonian for coupled oscillators given by

$$H = H_{cav} + H_o + W_{int}, \tag{2}$$

$$H = \hbar \begin{pmatrix} \omega_{cav} - j\kappa & 0 \\ 0 & 0 \end{pmatrix} + \hbar \begin{pmatrix} 0 & 0 \\ 0 & \omega_o - j\gamma \end{pmatrix} + \hbar \begin{pmatrix} 0 & \Omega/2 \\ \Omega/2 & 0 \end{pmatrix}, \tag{3}$$

$$\omega_{UP/LP} = \frac{\omega_{cav} + \omega_o}{2} - j\frac{\gamma + \kappa}{2} \\ \pm \sqrt{\Omega^2 + [(\omega_{cav} - \omega_o) + j(\gamma - \kappa)]^2}. \tag{4}$$

Here, $H_{cav}$ and $H_o$ are the bare Hamiltonians for respectively the cavity and magnon modes, and $W_{int}$ denotes their interaction. Correspondingly, $\omega_o$ and $\gamma$ are the magnon resonance frequency and linewidth, and $\omega_{cav}$ and

$\kappa$ are the cavity resonant frequency and linewidth, respectively. Finally, $\Omega$ is the splitting frequency and the only undetermined parameter in the equations. The solutions to this $2 \times 2$ Hamiltonian correspond to the upper and lower magnon-phonon-polaritons $\omega_{UP}$ and $\omega_{LP}$, respectively. As shown in Fig. 6b by the blue dashed curve, the coupled oscillator model (SC) shows excellent agreement with our data and indicates that the phonon, magnon, and cavity photons are no longer the eigenmodes of the system. Instead, the THz response goes back and forth at the frequency $\Omega$ between a state wherein the energy is stored purely in the light field and phonons in LiNbO$_3$ and a state wherein the energy stored purely in the magnons in ErFeO$_3$. Note that these spectra are distinctly different from the result of simple THz absorption by the magnon resonance in ErFeO$_3$. The magnon linewidth is far narrower than the splitting we observe between the upper and lower polariton peaks, as shown in Fig. 6b. Absorption at the magnon frequency would also yield a double-peaked structure around that frequency due to the magnon FID induced in the weak-coupling (WC) limit by the THz field generated in LiNbO$_3$. However, in this case the separation between the two peaks is given by the magnon absorption linewidth rather than the strength of the coupling (simulation shown in the red dotted curve of Fig. 6b). The actual spectrum we measure and the model based on strong coupling (blue dashed curve) show a far larger separation between the two THz spectral peaks.

By identifying the lower and higher frequency peaks as respectively the upper and lower polaritons, we generated a plot for the two branches as a function of the temperature, as shown in Fig. 7. The curves for the cou-pled oscillator modes and bare cavity and magnon modes are also overlaid in Fig. 7. The temperature dependence of the bare magnon frequency was recovered from the dips in the frequency spectrum (see Fig. 7, bottom left inset) and agrees well with previous results [47, 48]. The temperature dependence of the bare cavity frequencies was determined from an adjacent mode that showed the same dependence (see Fig. 7, top right inset). Subsequently, the bare frequencies were used to calculate the upper and lower polariton branches with $\Omega$ as the fitting parameter. The coupled oscillator model agrees well with the data at all detunings and reveals the existence of an avoided crossing between the two branches, a hallmark signature of strong coupling. From the two branches, we estimated the splitting frequency at zero detuning to be $\Omega/2\pi = 16$ GHz. Utilizing the splitting frequency and the linewidths for the magnon $\gamma/2\pi = 7$ GHz and cavity mode $\kappa/2\pi = 14$ GHz at zero detuning, we also estimated the cooperativity factor $C = \Omega^2/\kappa\gamma$, a dimensionless quantity that compares the coupling strength to the losses in the system. A system exhibiting $C > 1$ is considered to be in the strongly coupled regime, and we derived $C = 2.6$ for our data. This further supports the assertion that we reached the strong coupling regime.

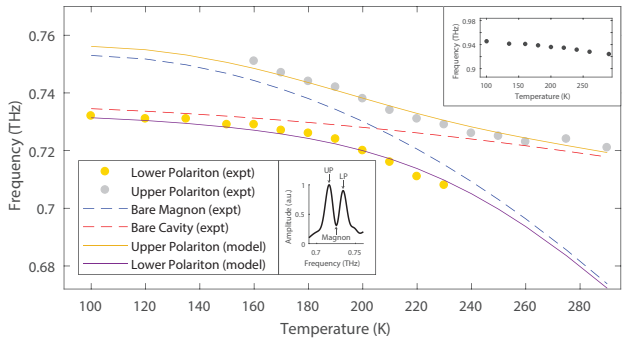

FIG. 7. **Avoided crossing of cavity magnon-phonon-polaritons.** The upper and lower polariton mode frequencies are plotted as a function of temperature. The solid lines correspond to the strongly coupled oscillator model while the dashed lines correspond to the bare magnon and cavity frequencies. The bottom left inset shows the locations of the upper (UP) and lower (LP) polariton and magnon frequencies in the cavity spectrum at 210 K. The top right inset shows the weakly quadratic temperature dependence of a higher-frequency cavity mode, which was used to determine the temperature dependence of the bare cavity mode that is resonant with the magnon (red dashed line).

Note that the quality factor of the hybrid cavity mode ranged from $Q = 42$ at room temperature to $Q = 56$ at 100 K. This was smaller than in our bare LiNbO$_3$ cavity and was attributed to both damping from the ErFeO$_3$ and significant radiative scattering. The radiative scattering is not entirely surprising given the difficulty of laser machining high quality air holes through a composite structure. Nonetheless, our Q was sufficient to reach the strong coupling regime, i.e. $C > 1$.

## IV. CONCLUSIONS

We have demonstrated evidence for strong coupling among magnons, phonons, and cavity photons at THz frequencies using a unique cavity platform. Firstly, we demonstrated a platform wherein THz waves are generated, confined, and detected directly inside a PhC cavity built into a LiNbO$_3$ slab. The cavity modes, strongly coupled admixtures of phonons and cavity photons, have a Q-factor of 510 at 79 K (the lowest temperature studied) and a mode volume of $V = 3.4 \times 10^{-3}\lambda^3 \simeq 0.5(\lambda/n)^3$ $\mu m^3$. The Q-factor is limited by the phononic nature of the modes and can be increased to $Q \simeq 1000$ at 4 K. Secondly, the platform was used to demonstrate strong coupling in a hybrid cavity composed of LiNbO$_3$ and ErFeO$_3$. Using temperature tuning of the cavity and magnon modes, we observed a normal mode splitting of 16 GHz, an avoided crossing between the upper and lower polariton branches, and a cooperativity factor of $C = 2.6$. The data also show excellent agreement with a coupled oscillator model. These observations are understood as evidence for the formation of cavity magnon-phonon-polaritons, wherein phonons in the LiNbO$_3$ and magnons in the ErFeO$_3$ are coupled to the EM wave E- and H-fields, respectively.

Our demonstration illustrates a system with multifunctional capabilities that should prove promising for realizing THz cavity spintronics. For instance, ErFeO$_3$ enables long-term storage and all-optical addressing of spin states. As a complement, LiNbO$_3$ has large electro-optic constants that enable efficient generation and detection of THz waves, with handles for both spatial and temporal shaping of the THz field profile.

Our cavity is able to achieve a Q-factor comparable to another THz cavity recently used in strong coupling work [49], but in a much smaller mode volume that is nearly diffraction limited. Another advantage is that the Q-factor can be increased without degrading the injection and readout. To inject and readout light from a typical photonic crystal cavity, one couples to free space via radiative scattering or directly couples into the mode via the evanescent field. Therefore, the Q-factor must be degraded to boost the coupling efficiency. In contrast, our platform circumvents this by generating and detecting the light directly inside the cavity, enabling us to increase the Q-factor without the same penalty. We believe this will prove valuable for experiments pursuing THz cavity QED. These experiments require high $Q$ and low $V$ because the vacuum Rabi frequency must be fast compared with the cavity loss rate and decoherence processes. A high $Q$ minimizes decoherence, while a lower $V$ increases the Rabi frequency ($\Omega \sim 1/\sqrt{V}$). There has already been some progress in using electro-optic crystals similar to LiNbO$_3$ to measure the quantum properties of light at THz frequencies [50–52]. Combin-ing this methodology with our cavity should be seamless, and should enable the demonstration of THz cavity QED phenomena in the future.

## ACKNOWLEDGMENTS

The authors would like to thank Professor Stanislav Kamba and Professor Shixun Cao for the ErFeO$_3$ sample. This work was supported by the Samsung Global Research Outreach program. A.S. acknowledges support via the Leopoldina Fellowship Programme, German National Academy of Sciences Leopoldina (LPDS 2016-02). P.S. acknowledges a Postgraduate Scholarship (PGS-D) from the Natural Sciences and Engineering Research Council of Canada (NSERC).

## Appendix A: Appendix

To calculate the energy distribution between the lattice motion and EM field, we begin by describing the coupling between the transverse polar optic phonon mode and the electromagnetic field. The coupled equations are given as [21].

$$P = \omega_{TO}\sqrt{\epsilon_o(\epsilon_0 - \epsilon_\infty)}Q + \epsilon_o(\epsilon_\infty - 1)E \quad (A1)$$

$$\ddot{Q} = -\omega_{TO}^2 Q - \Gamma\dot{Q} + \omega_{TO}\sqrt{\epsilon_o(\epsilon_0 - \epsilon_\infty)}E \quad (A2)$$

Here, $P$ is the macroscopic polarization of the material, $E$ is the E-electric field amplitude, and $Q$ is the normalized ionic displacement of the polar optic phonon mode where an over-dot indicates the time derivative. In addition, $\omega_{TO}$ is the transverse optic phonon frequency, $\epsilon_o$ is the vacuum permittivity, and $\varepsilon_\infty$ and $\varepsilon_0$ are the high- and low-frequency dielectric constants of the material, respectively. We now introduce Poynting's theorem, which relates the energy stored in the electromagnetic field to the work done on the electric dipoles in the medium. In differential form, the energy balance can be expressed as [53]

$$-\nabla \cdot S = \dot{W}, \quad (A3)$$

where $S$ is the Poynting vector and $W$ is the energy density. Equation (A3) states that the rate of energy decrease in the medium (RHS) is equal to the amount of energy flow out of the medium (LHS). To determine the energy distribution between the lattice and EM field, we must determine the appropriate energy density $W$ that satisfies Eqn. (A3). Such an energy density was proposed by Huang [21] in the absence of damping ($\Gamma = 0$), and is given as

$$W = \frac{1}{2}\left(\dot{Q}^2 + \omega_{TO}^2 Q^2\right) + \frac{1}{2}\left(\epsilon_o\varepsilon_\infty E^2 + \mu_o H^2\right) \quad (A4)$$

Where $H$ is the magnetic field and $\mu_o$ is the vacuum permeability. The first term in brackets represents the vibrational energy of the TO phonon mode, given as a sum of the potential and kinetic energy terms one normally uses for a harmonic oscillator. The second term in brackets represents the EM energy, with an inclusion of the electronic response via the term $\varepsilon_\infty$. From this energy density, we can write the time-averaged lattice and EM energies as

$$\langle W_{latt}\rangle = \frac{\omega_{TO}^2\epsilon_o(\epsilon_0 - \epsilon_\infty)\left(\omega^2 + \omega_{TO}^2\right)}{4\left(\omega_{TO}^2 - \omega^2\right)^2}|E_o|^2, \quad (A5)$$

$$\langle W_{EM}\rangle = \frac{1}{4}\left(\epsilon_o\varepsilon_\infty + \frac{1}{\mu_o}\left(\frac{k}{\omega}\right)^2\right)|E_o|^2. \quad (A6)$$

where $\langle W_{latt}\rangle$ and $\langle W_{EM}\rangle$ are the time-averaged energy of the lattice and EM wave, respectively. Using Eqns. (A5) and (A6), and assuming material parameters for LN as specified in [28], the fraction of energy in the lattice is 32% at 1 THz.

[1] P. Törmä and W. L. Barnes, Rep. Prog. Phys. **78**, 013901 (2015).

[2] M. Brune, F. Schmidt-Kaler, A. Maali, J. Dreyer, E. Hagley, J. M. Raimond, and S. Haroche, Phys. Rev. Lett. **76**, 1800 (1996).

[3] C. Weisbuch, M. Nishioka, A. Ishikawa, and Y. Arakawa, Phys. Rev. Lett. **69**, 3314 (1992).

[4] R. Ameling and H. Giessen, Nano Lett. **10**, 4394 (2010).

[5] Y. S. Park, A. K. Cook, and H. Wang, Nano Lett. **6**, 2075 (2006).

[6] H. Huebl, C. W. Zollitsch, J. Lotze, F. Hocke, M. Greifenstein, A. Marx, R. Gross, and S. T. B. Goennenwein, Phys. Rev. Lett. **111**, 127003 (2013).

[7] A. Wallraff, D. Schuster, A. Blais, L. Frunzio, R. Huang, J. Majer, S. Kumar, S. Girvin, and R. Schoelkopf, Nature **431**, 162 (2004).

[8] H. Walther, B. T. H. Varcoe, B.-G. Englert, and T. Becker, Reports Prog. Phys. **69**, 1325 (2006).

[9] J. M. Raimond, M. Brune, and S. Haroche, Rev. Mod. Phys. **73**, 565 (2001).

[10] T. D. Ladd, F. Jelezko, R. Laflamme, Y. Nakamura, C. Monroe, and J. L. O'Brien, Nature **464**, 45 (2010).

[11] A. Imamoglu, R. Ram, S. Pau, and Y. Yamamoto, Phys. Rev. A **53**, 4250 (1996).

[12] M. C. Cassidy, A. Bruno, S. Rubbert, M. Irfan, J. Kammhuber, R. N. Schouten, A. R. Akhmerov, and L. P. Kouwenhoven, Science (80-. ). **355**, 939 (2017).

[13] C. J. Hood, T. W. Lynn, A. C. Doherty, A. S. Parkins, and H. J. Kimble, Science (80-. ). **287**, 1447 (2000).

[14] M. a. Schmidt, D. Y. Lei, L. Wondraczek, V. Nazabal, and S. a. Maier, Nat. Commun. **3**, 1108 (2012).

[15] X. Zhang, C. L. Zou, L. Jiang, and H. X. Tang, Phys. Rev. Lett. **113**, 156401 (2014).

[16] R. W. Sanders, V. Jaccarino, and S. M. Rezende, Solid State Commun. **28**, 907 (1978).

[17] T. Kampfrath, A. Sell, G. Klatt, A. Pashkin, S. Mährlein, T. Dekorsy, M. Wolf, M. Fiebig, A. Leitenstorfer, and R. Huber, Nat. Photonics **5**, 31 (2011).

[18] a. V. Kimel, A. Kirilyuk, P. a. Usachev, R. V. Pisarev, a. M. Balbashov, and T. Rasing, Nature **435**, 655 (2005).

[19] S. Baierl, M. Hohenleutner, T. Kampfrath, A. K. Zvezdin, A. V. Kimel, R. Huber, and R. V. Mikhaylovskiy, Nat. Photonics **10**, 715 (2016).

[20] T. F. Nova, A. Cartella, A. Cantaluppi, M. Först, D. Bossini, R. V. Mikhaylovskiy, A. V. Kimel, R. Merlin, and A. Cavalleri, Nat. Phys. **1**, 1 (2016).

[21] K. Huang, in *Proc. R. Soc. London A Math. Phys. Eng. Sci.*, Vol. 208 (The Royal Society, 1951) pp. 352–365.

[22] K. Huang, Nature **167**, 779 (1951).

[23] D. L. Mills and E. Burstein, Reports Prog. Phys. **37**, 817 (2001).

[24] P. Sivarajah, J. Lu, M. Xiang, W. Ren, S. Kamba, S. Cao, and K. A. Nelson, arXiv Prepr. arXiv1611.01814 (2016).

[25] K. P. Cheung and D. H. Auston, Phys. Rev. Lett. **55**, 2152 (1985).

[26] P. C. M. Planken, L. D. Noordam, J. T. M. Kennis, and A. Lagendijk, Phys. Rev. B **45**, 7106 (1992).

[27] H. J. Bakker, S. Hunsche, and H. Kurz, Phys. Rev. B **50**, 914 (1994).

[28] T. Feurer, N. S. Stoyanov, D. W. Ward, J. C. Vaughan, E. R. Statz, and K. A. Nelson, Annu. Rev. Mater. Res. **37**, 317 (2007).

[29] K.-L. Yeh, M. C. Hoffmann, J. Hebling, and K. A. Nelson, Appl. Phys. Lett. **90**, 171121 (2007).

[30] C. Yang, Q. Wu, J. Xu, K. A. Nelson, and C. A. Werley, Opt. Express **18**, 26351 (2010).

[31] C. Winnewisser, P. U. Jepsen, M. Schall, V. Schyja, and H. Helm, Appl. Phys. Lett. **70**, 3069 (1997).

[32] I. P. Kaminow and W. D. Johnston Jr, Phys. Rev. **160**, 519 (1967).

[33] T. Feurer, J. C. Vaughan, and K. A. Nelson, Science **299**, 374 (2003).

[34] C. A. Werley, K. Fan, A. C. Strikwerda, S. M. Teo, X. Zhang, R. D. Averitt, and K. A. Nelson, Opt. Express **20**, 8551 (2012).

[35] P. Sivarajah, B. K. Ofori-Okai, S. M. Teo, C. A. Werley, and K. A. Nelson, New J. Phys. **17** (2015).

[36] E. Yablonovitch, Phys. Rev. Lett. **58**, 2059 (1987).

[37] S. John, Phys. Rev. Lett. **58**, 2486 (1987).

[38] J. D. Joannopoulos, S. G. Johnson, J. N. Winn, and R. D. Meade, *Photonic Crystals Molding the Flow of Light*, 2nd ed. (Princeton University Press, Princeton, Princeton, New Jersey, 2008) pp. 52–54.

[39] P. Sivarajah, C. A. Werley, B. K. Ofori-Okai, and K. A. Nelson, Appl. Phys. A **112**, 615 (2013).

[40] A. R. A. Chalcraft, S. Lam, D. O'Brien, T. F. Krauss, M. Sahin, D. Szymanski, D. Sanvitto, R. Oulton, M. S. Skolnick, A. M. Fox, D. M. Whittaker, H. Y. Liu, and M. Hopkinson, Appl. Phys. Lett. **90**, 241117 (2007).

[41] K. Wong, *Properties of Lithium Niobate* (The Institution of Engineering and Technology, 2002) p. 432.

[42] A. Ridah, M. D. Fontana, and P. Bourson, Phys. Rev. B **56**, 5967 (1997).

[43] T. Qiu and M. Maier, Phys. Rev. B **56**, R5717 (1997).

[44] H. J. Bakker, S. Hunsche, and H. Kurz, Rev. Mod. Phys. **70**, 523 (1998).

[45] L. Palfalvi, J. Hebling, J. Kuhl, A. Peter, and K. Polgar, J. Appl. Phys. **97**, 3505 (2005).

[46] N. Hodgson and H. Weber, *Laser Resonators and Beam Propagation: Fundamentals, Advanced Concepts, Applications*, Vol. 108 (Springer, 2005).

[47] G. V. Kozlov, S. P. Lebedev, A. A. Mukhin, A. S. Prokhorov, L. V. Fedorov, A. M. Balbashov, and I. Y. Parsegov, Magn. IEEE Trans. **29**, 3443 (1993).

[48] A. M. Balbashov, G. V. Kozlov, A. A. Mukhin, and A. S. Prokhorov, *Submillimeter spectroscopy of antiferromagnetic dielectrics. Rare-earth orthoferrites* (World Scientific Pub. Co. Inc, 1995).

[49] Q. Zhang, M. Lou, X. Li, J. L. Reno, W. Pan, J. D. Watson, M. J. Manfra, and J. Kono, Nat. Phys. **1**, 1 (2016).

[50] C. Riek, D. V. Seletskiy, a. S. Moskalenko, J. F. Schmidt, P. Krauspe, S. Eckart, S. Eggert, G. Burkard, and A. Leitenstorfer, Science (80-. ). **350**, 420 (2015).

[51] A. S. Moskalenko, C. Riek, D. V. Seletskiy, G. Burkard, and A. Leitenstorfer, Phys. Rev. Lett. **115**, 263601 (2015).

[52] C. Riek, P. Sulzer, M. Seeger, A. S. Moskalenko, G. Burkard, D. V. Seletskiy, and A. Leitenstorfer, Nature **541**, 376 (2017).

[53] D. J. Griffiths and C. Inglefield, Am. J. Phys. **73**, 574 (2005).